\documentclass[reprint,superscriptaddress,amsmath,amssymb,aps,prl,notitlepage,floatfix,nofootinbib,longbibliography,twocolumn]{revtex4-1}

\usepackage{changes}
\usepackage{tensor}     
\usepackage{graphicx}   
\usepackage[
colorlinks=true,        
citecolor=blue,         
linkcolor=blue,         
urlcolor=blue           
]{hyperref}             
\usepackage{bm}         
\usepackage{xcolor}     
\usepackage{lipsum}
\usepackage{color}      
\usepackage[utf8]{inputenc} 
\usepackage[section]{placeins} 
\usepackage{multirow}
\usepackage{orcidlink}
\newcommand{\nc}{\newcommand*} 

\nc{\al}{\alpha}
\nc{\s}{\sigma}
\nc{\kp}{\kappa}
\nc{\dt}{\delta}
\nc{\Dt}{\Delta}
\nc{\Ld}{\Lambda}
\nc{\p}{\partial}
\nc{\Gm}{\Gamma}
\nc{\om}{\omega}
\nc{\Om}{\Omega}
\nc{\rd}{\mathrm{d}}
\def\({\left(}
\def\){\right)}
\def\[{\left[}
\def\]{\right]}
\def\e{\begin{equation}}
\def\q{\end{equation}}
\def\m{\begin{eqnarray}}
\def\n{\end{eqnarray}}
\nc{\Eq}[1]{Eq.~\eqref{#1}}     
\nc{\Eqs}[1]{\eqref{#1}} 
\nc{\Fig}[1]{Fig.~\ref{#1}}     
\nc{\Table}[1]{Table~\ref{#1}}  
\nc{\Sec}[1]{Sec.~\ref{#1}}     
\nc{\Msun}{M_\odot}             
\nc{\fpbh}{f_{\mathrm{pbh}}}    
\nc{\fpbhn}{f_{\mathrm{pbh0}}}    
\nc{\mR}{\mathcal{R}} 
\nc{\seq}{\sigma_{\mathrm{eq}}}
\nc{\ogw}{\Omega_{\mathrm{GW}}}
\nc{\gpcyr}{\mathrm{Gpc}^{-3}\,\mathrm{yr}^{-1}}
\nc{\lvc}{LIGO/Virgo} 
\nc{\SNR}{\mathrm{SNR}} 
\nc{\mmin}{{m_{\mathrm{min}}}}
\nc{\mmax}{{m_{\mathrm{max}}}}
\nc{\Mmin}{{M_{\mathrm{min}}}}
\nc{\fmin}{{f_{\mathrm{min}}}}
\nc{\VT}{\mathrm{VT}}
\nc{\rhoGW}{\rho_{\mathrm{GW}}}
\nc{\vth}{\vec{\theta}}
\nc{\vd}{\vec{d}}
\nc{\vla}{\vec{\lambda}}
\nc{\Nobs}{N_{\mathrm{obs}}}
\nc{\av}[1]{\langle #1 \rangle} 
\nc{\km}{\mathrm{km}}
\nc{\Mpc}{\mathrm{Mpc}}
\nc{\Tobs}{T_{\mathrm{obs}}}
\nc{\Ntemp}{N_{\mathrm{temp}}}
\nc{\fyr}{f_{\mathrm{yr}}}
\nc{\addref}{[\textcolor{red}{add ref}] } 
\nc{\eg}{\textit{e.g.~}}
\nc{\app}{\approx}
\nc{\hf}{\frac{1}{2}}
\nc{\discuss}{\textcolor{red}{Add discussion here!}}
\nc{\red}[1]{\textcolor{red}{#1}}
\nc{\hp}{h_+} 
\nc{\hc}{h_{\times}} 
\nc{\Oh}{\hat{\Omega}}
\nc{\vx}{\vec{x}}
\nc{\mh}{\hat{m}}
\nc{\nh}{\hat{n}}
\nc{\zh}{\hat{z}}
\nc{\ph}{\hat{p}}
\nc{\A}[1]{\mathcal{A}_{#1}}
\nc{\Ogw}[1]{\Omega_{\mathrm{#1}}}
\nc{\bn}[1]{\dt{t}_{\text{#1}}}
\nc{\bC}[1]{{C}_{\text{#1}}}
\nc{\NTOA}{N_{\text{TOA}}}
\nc{\Nmode}{{N_{\text{mode}}}}
\nc{\ARN}{A_{\rm{RN}}}
\nc{\gRN}{\gamma_{\rm{RN}}}
\nc{\bS}{{\Sigma}}
\nc{\br}{{r}}
\nc{\bN}{{R}}
\nc{\Agw}{A_\mathrm{GWB}}
\nc{\UCP}{\mathrm{UCP}}
\nc{\TT}{\mathrm{TT}}
\nc{\ST}{\mathrm{ST}}
\nc{\SL}{\mathrm{SL}}
\nc{\VL}{\mathrm{VL}}
\nc{\lpr}{l^{\prime}}
\nc{\mpr}{m^{\prime}}
\nc{\Cth}{\mathcal{C}_{\mathrm{th}}}

\nc{\BFST}{$107 \pm 7$}
\begin{document}
	
\title{Constraints on the Primordial Black Hole Abundance using Pulsar Parameter Drifts}

\author{Yan-Chen Bi\orcidlink{0000-0002-9346-8715}}
\affiliation{Institute of Theoretical Physics, Chinese Academy of Sciences, Beijing 100190, China}
\affiliation{School of Physical Sciences, University of Chinese Academy of Sciences, No. 19A Yuquan Road, Beijing 100049, China}

\author{Yu-Mei Wu\orcidlink{0000-0002-9247-5155}}
\email{Corresponding author: wuyumei@yzu.edu.cn} 
\affiliation{Center for Gravitation and Cosmology, College of Physical Science and Technology, Yangzhou University, Yangzhou, 225009, China}

\author{Qing-Guo Huang\orcidlink{0000-0003-1584-345X}}
\email{Corresponding author: huangqg@itp.ac.cn}
\affiliation{Institute of Theoretical Physics, Chinese Academy of Sciences, Beijing 100190, China}
\affiliation{School of Physical Sciences, 
    University of Chinese Academy of Sciences, 
    No. 19A Yuquan Road, Beijing 100049, China}
\affiliation{School of Fundamental Physics and Mathematical Sciences, Hangzhou Institute for Advanced Study, UCAS, Hangzhou 310024, China}

\begin{abstract}
Primordial black holes (PBHs) provide a compelling interpretation for the binary black holes (BBHs) observed by ground-based gravitational-wave (GW) detectors, especially for those BBHs in the theoretical mass gap. 
In the early Universe, the scalar perturbations required to produce such PBHs inevitably generate scalar-induced GWs (SIGWs).
These SIGWs peak in the sub-nanohertz band, and manifest secularly as measurable jerk-like drifts in the second derivative of pulsar spin periods. 
In this Letter, we perform the first search for SIGWs using pulsar parameter drifts, and place a 95\% confidence-level upper limit on the PBH abundance of $f_{\mathrm{PBH}} < 10^{-10}$ over the mass range $[3 \times 10^{-1}, 4 \times 10^{4}]\Msun$. Our results strongly disfavor a PBH origin for the BBHs currently detected by the LIGO–Virgo–KAGRA (LVK) Collaborations.
\end{abstract}

\maketitle


\textit{Introduction.} 
\label{sec:introduction}
Since the first detection of gravitational waves (GWs), ground-based detectors such as the Laser Interferometer Gravitational-Wave Observatory (LIGO) have cataloged hundreds of binary black hole (BBH) mergers spanning a broad mass distribution \cite{LIGOScientific:2018mvr,LIGOScientific:2020ibl,LIGOScientific:2021usb,KAGRA:2021vkt,LIGOScientific:2025slb}. Stellar evolution, however, predicts a dearth of black holes (BHs) in the approximate mass range $45-130\Msun$, known as the upper BH mass gap, because very massive stars can reach a stage where a pair-instability supernova (PISN) occurs. There, the creation of electron-positron pairs temporarily lowers the radiation pressure supporting the stellar core, and triggers a contraction that can ignite a runaway thermonuclear explosion \cite{1967ApJ...148..803R}. Such process may completely disrupt the progenitor and leave no compact remnant \cite{Fraley1968}, strongly suppressing BH formation in this mass range \cite{Barkat:1967zz,2019ApJ...887...53F,Marchant:2018kun}. Nevertheless, a particularly intriguing subset of BBH detections appears in this mass gap, including GW190426\_190642 \cite{LIGOScientific:2021usb}, GW190521 \cite{LIGOScientific:2020iuh}, and the more recent GW231123 \cite{LIGOScientific:2025rsn}. Their existence therefore points to formation channels beyond ordinary stellar collapse and motivates non-stellar interpretations of the BBH population observed by the LIGO-Virgo-KAGRA (LVK) Collaboration.

Primordial BHs (PBHs), seeded by the catastrophic gravitational collapse of primordial density fluctuations in the early Universe \cite{Hawking:1971ei,Carr:1974nx,Chapline:1975ojl}, offer a natural and compelling resolution \cite{Bird:2016dcv,Sasaki:2016jop,Carr:2019kxo,Clesse:2020ghq,DeLuca:2020sae,Hutsi:2020sol,Wong:2020yig,Jedamzik:2020omx,DeLuca:2021wjr,Chen:2021nxo,Huang:2024wse,Yuan:2024yyo,Yuan:2025avq,Chen2025,Magaraggia:2026jhk}. Beyond readily circumventing the PISN mass gap, PBHs remain premier candidates for cold dark matter (CDM) and vital seeds for early galaxy formation \cite{Carr:2016drx,Bean:2002kx,Kawasaki:2012kn}. Since their theoretical inception, the abundance of PBHs has been subjected to a broad and increasingly stringent set of multi-messenger constraints (see \cite{Carr:2020gox} for a comprehensive review). These include searches for subsolar-mass BBHs and GW backgrounds (GWBs) from LIGO \cite{Wang:2016ana,Kavanagh:2018ggo,LIGOScientific:2019kan,Nitz:2022ltl}, the null detection of a GWB by pulsar timing arrays (PTAs) \cite{Chen:2019xse}, microlensing searches by Optical Gravitational Lensing Experiment (OGLE) \cite{Mroz:2024wia,Mroz:2024mse}, radio and  X-ray limits from gas accretion onto PBHs \cite{Gaggero:2016dpq,Manshanden:2018tze}, dynamical-friction (DF) bounds from spheroid stars and halo objects \cite{Carr:2020gox}, and CMB constraints on accretion-induced distortions \cite{Serpico:2020ehh}.

Crucially, the extreme cosmological scalar perturbations required to form PBHs inevitably source scalar-induced GWs (SIGWs) upon horizon re-entry \cite{Ananda:2006af,Baumann:2007zm,Saito:2008jc}, thereby enabling direct constraints on PBHs \cite{Saito:2009jt,Bugaev:2010bb,Clesse:2018ogk,Chen:2019xse,Romero-Rodriguez:2021aws,Jiang:2024aju}. We refer readers to reviews and the references therein \cite{Yuan:2021qgz,Domenech:2021ztg}. The characteristic peak frequency of this secondary emission, is inherently linked to the PBH mass via $f_{\star} \sim 3\mathrm{nHz} (m_{\mathrm{PBH}}/\Msun)^{-1/2}$ \cite{Saito:2008jc}. Consequently, the formation of LIGO-band PBHs with $\mathcal{O}(1-100) \Msun$ generically produces SIGWs deep within the sub-nanohertz regime. However, although PTAs offer the closest observational probe of such low frequencies, these signals nevertheless remain below their nominal sensitivity window, as GW power beneath the fundamental frequency resolution, $1/T_{\mathrm{obs}} \sim 10^{-9}\mathrm{Hz}$, where $T_{\mathrm{obs}}$ is the observational timespan, is absorbed into pulsar timing models and removed from post-fit timing residuals \cite{Hazboun:2019vhv,Pitrou:2024scp,Allen:2025waa}. Yet, while absent from the post-fit timing residuals, this SIGW signature is not destroyed, but instead encoded in the best-fit pulsar timing parameters, accumulating over decades of observation and manifesting as measurable secular drifts \cite{DeRocco:2022irl,DeRocco:2023qae,Zheng:2025tcm}. 

In this Letter, we exploit secular pulsar parameter drifts to conduct the first search for SIGW signals in the sub-nHz frequency range. By bypassing the low-frequency cutoff of standard PTA residual analyses, we place the strongest constraints to date on the PBH abundance across the critical mass range of $[3 \times 10^{-1}, 4 \times 10^4]\Msun$.

\textit{SIGWs search using pulsar parameter drifts.} 
\label{sec:sigw}
The energy density of GWs per logarithmic frequency interval, $[f,f+d\ln f]$, normalized by the critical density $\rho_c$ that closes the Universe, is defined as
\m
\Ogw{GW}(f) = \frac{1}{\rho_c} \frac{d \rho_{\mathrm{GW}}}{d \ln f} .
\n
For the SIGWs produced during radiation domination, the leading contribution to the SIGW energy density fraction is given by \cite{Kohri:2018awv,Espinosa:2018eve}
\e
\begin{aligned}
\Ogw{GW}(k) \! =\! c_g \Omega_{\mathrm{r},0} &\int_0^1 dx \int_1^{\infty} dy \mathcal{F}(x,y) \\
& \times \mathcal{P}_{\zeta}(k\frac{y-x}{2}) \mathcal{P}_{\zeta}(k\frac{y+x}{2}) .
\end{aligned}
\q
Here, $\Omega_{\mathrm{r},0} = 9 \times 10^{-5}$ is the present radiation energy-density fraction, $\mathcal{P}_{\zeta}(k)$ is the dimensionless curvature power spectrum, $\mathcal{F}(x,y)$ is the kernel function, and $c_g$ accounts for the change in relativistic degrees of freedom between GW production and today. The curvature perturbation satisfies $\langle \zeta(\vec{k}) \zeta(\vec{k}^{\prime}) \rangle = (2\pi^2 / k^3) \mathcal{P}_{\zeta}(k)\, \delta(\vec{k}+\vec{k}^{\prime})$, 
where $\delta$ is the Dirac delta function and $k$ is the comoving wavenumber relating to the present frequency as $f = k / (2\pi a_0)$ with $a_0$ the present scale factor. The factor $c_g$ is given by $c_g = (g_{\star,\mathrm{rad}} / g_{\star,0}) (g_{\star S,0} / g_{\star S,\mathrm{rad}})^{4/3}$, where $g_{\star}$ and $g_{\star S}$ denote the effective numbers of degrees of freedom for the energy and entropy densities, respectively, evaluated at the GW production epoch (``rad'') and today ( ``0''). For Standard Model particle content with all species relativistic at production, one finds $c_g \sim 0.4$ \cite{Espinosa:2018eve}. The kernel $\mathcal{F}(x,y)$ takes the form 
\begin{widetext}
\m
\mathcal{F}(x,y) = \frac{12 (x^2 + y^2 - 6)^4 (x^2-1)^2 (y^2-1)^2}{(x-y)^8 (x+y)^8}
\left[\left(\frac{2(x^2 - y^2)}{x^2 + y^2 - 6} + \ln \left\vert \frac{y^2 - 3}{x^2 - 3}\right\vert\right)^2 + \pi^2 \Theta(y - \sqrt{3})\right],
\n
\end{widetext}
where $\Theta(x)$ denotes the Heaviside step function.

Both the amplitude and shape of $\Ogw{GW}(k)$ 
are determined by the scalar power spectrum $\mathcal{P}_{\zeta}(k)$. A Dirac $\delta$-function spectrum peaked at $k_\star$,
$\mathcal{P}_{\zeta}(k) = A\delta\[\ln(k/k_\star)\]$, corresponds to a monochromatic PBH formation. However, in literature, a more realistic spectrum with log-normal shape is widely adopted, 
\m
\mathcal{P}_{\zeta}(k) = \frac{A}{\sqrt{2\pi}\Delta} \exp \(- \frac{\ln^2(k/k_{\star})}{2 \Delta^2} \) ,
\label{lognormal}
\n
where $A$ and $k_{\star}$ denote the amplitude and peak position of the power spectrum, respectively, and $\Delta$ determines the width of the spectrum. Such a spectrum can arise naturally in many inflation models, and reduces to the Dirac $\delta$-function power spectrum in the limit of $\Delta\rightarrow 0$. 
For $\Delta \ll 1$, both the amplitude of SIGWs and the generated PBH abundance become independent of $\Delta$ \cite{Romero-Rodriguez:2021aws}. 
Throughout this work we adopt $\Delta = 0.1$ as a benchmark choice for a quasi-monochromatic PBH formation.

When pulsars are used as probes, ultra-low-frequency GWs ($f \le 1\mathrm{nHz}$) do not appear in timing residuals but instead induce secular shifts in pulsar spin parameters.
Specifically, these SIGWs, which peak in the sub-nanohertz band, induce a jerk-like signal, $j_{\mathrm{GW}}$, which is effectively absorbed into the best-fit second derivative of the spin period, $\ddot{P}_{\mathrm{obs}}$, during the standard timing model fit.
Importantly, standard kinematic and Galactic contaminations, such as the Shklovskii effect and Galactic gravitational acceleration, are orders of magnitude below current experimental sensitivities, rendering $\ddot{P}_{\mathrm{obs}}$ a clean probe of these SIGWs \cite{DeRocco:2022irl}.

Assuming that these sub-nanohertz signals are fully absorbed into the timing-model parameters \cite{DeRocco:2023qae}, with the GW metric perturbations $h_{A}(f,\hat{\Omega})$, characterized by polarization $A=+,\times$ and propagation direction $\hat{\Omega}$, following a zero-mean Gaussian distribution with covariance
\m
\langle h_A^{\star}(f,\hat{\Omega}) h_{A^{\prime}}(f^{\prime},\hat{\Omega}^{\prime}) \rangle 
= \frac{S_h(f)}{2} \delta(f-f^{\prime}) \frac{\delta(\hat{\Omega}, \hat{\Omega}^{\prime})}{4\pi} \delta_{AA^{\prime}} ,
\n
the two-point cross-correlation of the induced jerk between a pair of pulsars $a$ and $b$ is given by
\e
\begin{aligned}
\langle j^{(a)}_{\mathrm{GW}} j^{(b)}_{\mathrm{GW}} \rangle = \int_0^{f_T} S_h(f) (2\pi f)^4 \mathrm{Re} \Gamma(\gamma_{ab},f) df .
\end{aligned}
\label{correlator}
\q
Here $S_h(f)$ denotes the GWB power spectrum, related to $\ogw$ via $S_h(f) = 3H_0^2\ogw/(4\pi^2 f^3)$, where $H_0 = 100\,h_0\, \mathrm{km} \ \mathrm{s^{-1} Mpc^{-1}}$ and $h_0 \approx 0.67$ \cite{Planck:2018vyg}. The integration is truncated at $f_T=1/(4\,T_{\mathrm{max}})$ to ensure the validity of the Taylor expansion of the timing model around $t=0$, where $T_{\mathrm{max}}$ is the maximum observation baseline \cite{DeRocco:2022irl,DeRocco:2023qae}. The geometric response of the pulsar pair, separated by an angle $\gamma_{ab}$, is encoded in 
\m
\Gamma(\gamma_{ab},f) = \int \frac{d\hat{\Omega}}{4\pi} \sum_{A} R_a^A(f,\hat{\Omega}) R_b^A(f,\hat{\Omega}),
\n
where the antenna pattern function, including the pulsar term, is
\m
R_a^A(f,\hat{\Omega}) = \left[1 - \mathrm{e}^{-i2\pi f d_a(1 + \hat{\Omega} \cdot \hat{p}_a)}\right]
\frac{\hat{p}_a^i \hat{p}_a^j \hat{e}_{ij}^A(\hat{\Omega})}{2 (1 + \hat{\Omega} \cdot \hat{p}_a)} ,
\n
with $\hat{e}_{ij}^A$ the polarization tensor, and $d_a$, $\hat{p}_a$ the distance and direction of pulsar $a$.

To constrain the SIGW amplitude, we further develop the effective framework introduced in \cite{DeRocco:2023qae}.
We define the observable for each pulsar as
\m
j_{\mathrm{obs}} \equiv \frac{\ddot{P}_{\mathrm{obs}}}{P},
\n
which receives contributions from both noise and the GW signal,
\m
j_{\mathrm{obs}} = n_{\mathrm{obs}} + n_{\mathrm{RN}} + j_{\mathrm{GW}},
\label{j_obs}
\n
where $n_{\mathrm{obs}}$ and $n_{\mathrm{RN}}$ denote the measurement uncertainty and ultralow-frequency red noise (RN), respectively.
The data vector $\mathbf{j}_{\mathrm{obs}}$ of length of pulsar number $N_p$ is modeled with a Gaussian likelihood,
\m
\mathcal{L} = \frac{1}{\sqrt{(2\pi)^{N_p} |\mathbf{C}|}} 
\exp\left[-\frac{1}{2}\mathbf{j}_{\mathrm{obs}}^{T} \mathbf{C}^{-1} \mathbf{j}_{\mathrm{obs}}\right],
\label{likelihood}
\n
with covariance matrix $\mathbf{C} = \mathbf{N}_{\mathrm{obs}} + \mathbf{N}_{\mathrm{RN}} + \mathbf{C}_{\mathrm{GW}}$, corresponding to the decomposition in Eq.~(\ref{j_obs}).
Specifically, the diagonal measurement-noise component, $N_{\mathrm{obs},ab} = \delta_{ab} \sigma^2_{\mathrm{obs}}$, includes contributions from dispersion-measure (DM) variations and high-frequency red noise \cite{Liu:2019iuh}. The ultralow-frequency RN component is given by $N_{\mathrm{RN},ab} = \delta_{ab} \sigma^2_{\mathrm{RN}} = \delta_{ab} \int_0^{f_T} df \, (2\pi f)^4 S_{\mathrm{RN}}(f)$, where the RN spectrum is modeled as a broken power law, $S_{\mathrm{RN}}(f) = \frac{A_{\mathrm{RN}}^2 f^2}{3 f_{\mathrm{yr}}^3}
\left[1 + (f/f_c)^2\right]^{-\gamma_{\mathrm{RN}}/2}$, 
with $f_{\mathrm{yr}} = 1\,\mathrm{yr}^{-1}$ the reference frequency. The parameters $\gamma_{\mathrm{RN}}, f_c,$ and $A_{\mathrm{RN}}$ are marginalized over using priors from the EPTA and PPTA analyses. The resulting values of $\sigma^2_{\mathrm{obs}}$ and $\sigma^2_{\mathrm{RN}}$ are given in the Supplemental Material (SM). Finally, the GW-induced covariance $C_{\mathrm{GW},ab} = \langle j^{(a)}_{\mathrm{GW}} j^{(b)}_{\mathrm{GW}} \rangle$, as specified by Eq.~(\ref{correlator}), encodes the spatial correlations between pulsars induced by the stochastic GW background.

To set an upper limit on the SIGW amplitude, we evaluate the log-likelihood ratio between 
a signal model, specified by an overall normalization $S_0$, and the null hypothesis of no GW signal ($S_0 = 0$),
\m
\hat{r}(S_0) = -2 \log \left( \frac{\mathcal{L}(S_0 \mid \mathbf{j}_{\mathrm{obs}})}{\mathcal{L}(S_0 = 0 \mid \mathbf{j}_{\mathrm{obs}})} \right) .
\label{by}
\n
Following standard thresholding practices \cite{Algeri:2019lah,DeRocco:2023qae}, we establish the 95\% confidence limit on $S_0$ by requiring $\hat{r}(S_0)=2.71$.

\begin{figure}[tpb]
\centering
\includegraphics[width=1.\linewidth]{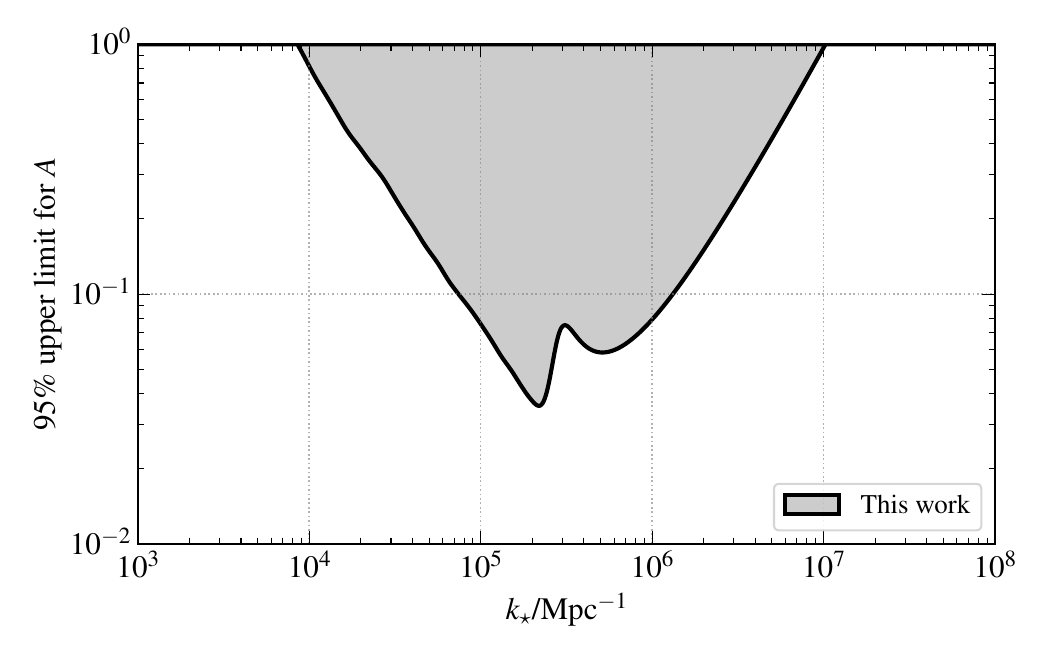}
\caption{ \label{fig:upper_a} The 95\% upper limits on the power spectrum amplitude A of scalar perturbation as a function of $k_{\star}$ using pulsar parameter drifts.}
\end{figure}

We apply this procedure to the pulsar parameter drift dataset for $\ddot{P}_{\mathrm{obs}}$ constructed in \cite{DeRocco:2023qae}, which is based on timing data from the EPTA \cite{Reardon:2015kba} and PPTA \cite{EPTA:2016ndq} collaborations originally processed in \cite{Liu:2019iuh}. The dataset is listed in SM for reference. Evaluating the likelihood ratio defined by \Eq{likelihood} and \Eq{by}, we obtain 95\% upper limits on the primordial scalar power-spectrum amplitude $A$ as a function of the peak position $k_{\star}$, shown in \Fig{fig:upper_a}. We find $A \lesssim 1$ throughout the range $[9 \times 10^{3}, 1 \times 10^{7}]$Mpc$^{-1}$. We have verified that moderate variations in the measurement uncertainties do not change the qualitative conclusion.


\textit{Constraints on PBH abundance.}
\label{sec:pbh}
Enhanced primordial scalar perturbations $\zeta(\vec{k})$ re-entering the Hubble horizon during the radiation-dominated era not only source SIGWs, but can also overcome radiation pressure to undergo direct gravitational collapse and form PBHs. This connection allows the constraints on the SIGW spectrum derived above to be ultimately related to PBH formation.

The primary phenomenological quantity bridging this early-universe mechanism to present cosmological observables is the PBH abundance, $f_{\mathrm{PBH}}$, which quantifies the fraction of cold dark matter (CDM) comprised of PBHs today
\m
f_{\mathrm{PBH}} = \frac{\Omega_{\mathrm{PBH}}}{\Omega_{\mathrm{CDM}}} = \int \psi_{\mathrm{PBH}}(m) d\ln m .
\n
Here, $\Omega_{\mathrm{CDM}} = 0.264$ is the present CDM density of the universe \cite{Planck:2018vyg}, and $\psi_{\mathrm{PBH}}(m)$ denotes the PBH mass function within the logarithmic PBH mass interval $[m,m + d \ln m]$, expressed as
\m
\psi_{\mathrm{PBH}}(m) = \frac{1}{\Omega_{\mathrm{CDM}}} \(\frac{M_{\mathrm{eq}}}{m}\)^{1/2} \beta,
\n
where $M_{\mathrm{eq}} \simeq 2.8 \times 10^{17} \Msun$ represents the horizon mass at matter-radiation equality, 
and $\beta$ is the initial PBH formation fraction evaluated at the corresponding mass scale. Its evaluation relies on the collapse criterion formulated in terms of the compaction function $\mathcal{C}$, defined as the local mass excess relative to the background averaged over a given comoving radius \cite{Ianniccari:2024bkh}. More explicitly, $\beta$ is obtained by integrating the joint probability distribution function of the compaction function \cite{Jiang:2024aju,Ianniccari:2024bkh}
\m
\beta = \int_{\mathcal{D}} d\mathcal{C}_l(r_m) d\mathcal{C}_l^{\prime\prime}(r_m) \frac{m}{M_H} P(\mathcal{C}_l(r_m), \mathcal{C}_l^{\prime\prime}(r_m)) ,
\n
where the domain of integration is defined by $\mathcal{D}=\{\mathcal{C}>\Cth \wedge \mathcal{C}_l < 4/3\}$. Here, $\mathcal{C}_l$ denotes the linear counterpart of $\mathcal{C}$, related via $\mathcal{C} = \mathcal{C}_l - (3/8)\mathcal{C}_l^2$, and $\Cth$ is the critical threshold for collapse. Following \cite{Musco:2020jjb,Escriva:2019phb}, we take $\Cth \approx 0.58$, with the characteristic scale defined by the radius $r_m$ at which $\mathcal{C}$ is maximized.
Noted that we only consider type I PBHs throughout this Letter. The PBH mass is indicated by the critical collapse scaling relation 
\m
m = M_H \kappa (\mathcal{C}(r_m) - \Cth(r_m))^{\gamma_c},
\n
where $\kappa= 3.3$, $\gamma_c = 0.36$ \cite{Koike:1995jm}, and $M_H \simeq 1.4 \times 10^{13} \Msun \({k}/{\Mpc^{-1}}\)^{-2}$ is the horizon mass at the time of PBH formation. 
The joint probability density function follows a Gaussian distribution \cite{Ianniccari:2024bkh}
\e
\begin{aligned}
P(\mathcal{C}_l(r_m), & \mathcal{C}_l^{\prime\prime}(r_m)) = \frac{1}{2\pi \sigma_0 \sigma_2 \sqrt{1-\gamma^2}} \exp\[ - \frac{r_m^4 \mathcal{C}_l^{\prime\prime}(r_m)^2}{32 \sigma_2^2} \] \\
&\times \exp\[-\frac{1}{2(1 - \gamma^2)} \(\frac{\mathcal{C}_l(r_m)}{\sigma_0} + \gamma \frac{r_m^2 \mathcal{C}_l^{\prime\prime}(r_m)}{4 \sigma_2}\)^2\] .
\end{aligned}
\q
The associated variance parameters are given by
\m
\sigma_0^2 &=& \langle \mathcal{C}_l(r_m)^2 \rangle, \\
\sigma_1^2 &=& -\frac{1}{4} r_m^2 \langle \mathcal{C}_l^{\prime\prime}(r_m) \mathcal{C}_l(r_m) \rangle,\\
\sigma_2^2 &=& \frac{1}{16} r_m^4 \langle \mathcal{C}_l^{\prime\prime}(r_m)^2 \rangle ,
\n
with $\gamma=\sigma_1^2 / (\sigma_2 \sigma_0)$.
These quantities can be expressed in terms of the primordial scalar perturbation $\zeta(\vec{k})$ in Fourier space, through the linear compaction function
\m
\mathcal{C}_l(\vec{k},r) = \frac{4}{9} (kr)^2 W(k,r) \zeta(\vec{k}) \mathcal{T}^2(k,r) .
\n
Here, the smoothing function and sub-horizon evolution are encoded in the top-hat window function $W(k,r)$ and the radiation transfer function $\mathcal{T}(k,r)$ (assuming pure radiation domination) \cite{Byrnes:2025tji}, given respectively by
\m
W(k, r) = 3 \frac{\sin(kr) - (kr)\cos(kr)}{(kr)^3} ,    
\n
and
\m
\mathcal{T}(k, r) \!=\! 3 \frac{\sin(kr/\sqrt{3}) - (kr/\sqrt{3})\cos(kr/\sqrt{3})}{(kr/\sqrt{3})^3} .
\label{transfer}
\n

\begin{figure}[htpb]
\centering
\includegraphics[width=1.\linewidth]{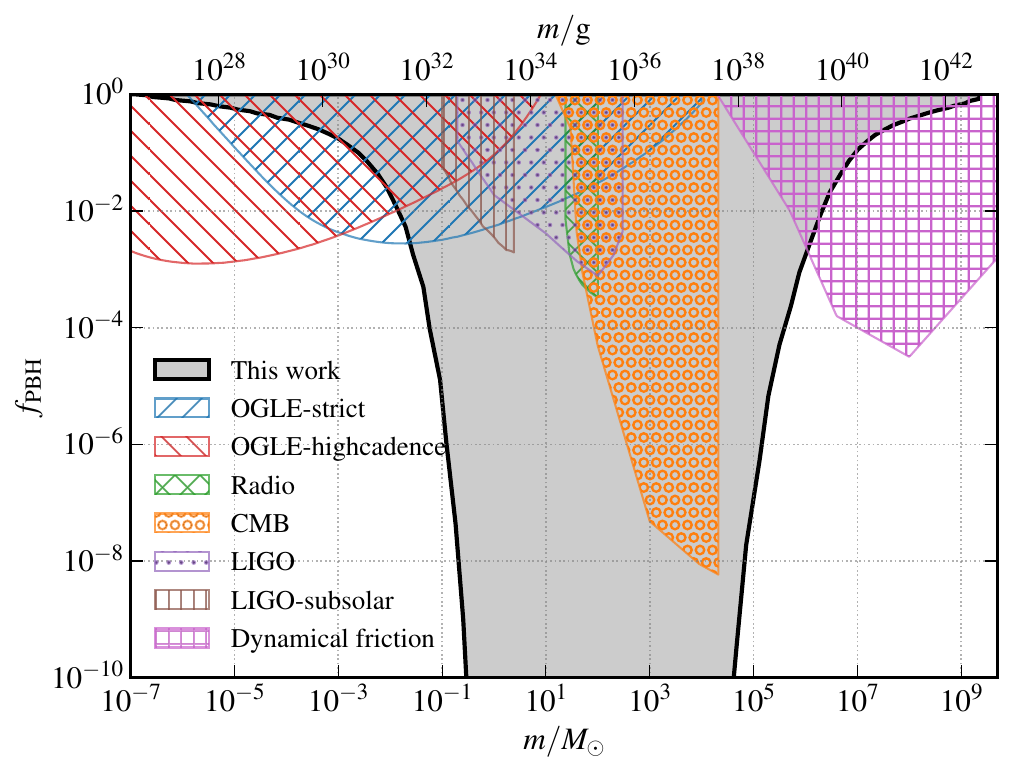}
\caption{ \label{fig:fpbh} Constraints on the abundance of PBHs in DM $\fpbh$ as a function of the PBH mass $m$ using pulsar parameter drifts. Results from OGLE \cite{Mroz:2024wia,Mroz:2024mse}, Radio \cite{Manshanden:2018tze}, CMB disk accretion \cite{Serpico:2020ehh}, LIGO \cite{Nitz:2022ltl,LIGOScientific:2019kan,Kavanagh:2018ggo} and Dynamical friction (DF) \cite{Carr:2020gox} are also shown.}
\end{figure}

The above formalism establishes a direct mapping from $\zeta(k)$ to the present-day PBH abundance, enabling the constraints on SIGWs to be translated into limits on $\fpbh$. The resulting 95\% confidence-level upper limits are shown in \Fig{fig:fpbh} as a function of the PBH mass $m$. We find that secular pulsar parameter drifts constrain $\fpbh < 10^{-10}$ over the mass range $[3 \times 10^{-1}, 4 \times 10^{4}]\,\Msun$.



\textit{Conclusion and Discussion.} 
In this Letter, we have presented the first search for the sub-nanohertz SIGWs that inevitably accompany the formation of stellar-mass PBHs. By pivoting away from standard timing residuals and instead examining the secular imprints left on the best-fit pulsar spin parameters $\ddot{P}_{\mathrm{obs}}$, we bypass conventional low-frequency cutoffs. Utilizing data from the EPTA and PPTA \cite{DeRocco:2023qae}, we establish the strongest bounds to date on the PBH abundance within the LVK BH mass window. 

The physical implications of our results are significant. Our constrained mass range spans the mass distribution of BBHs detected by the LVK Collaboration; interpreting these mergers as primordial origin would require an abundance of $f_{\mathrm{PBH}} \sim 0.01$ \cite{Vaskonen:2019jpv,Vaskonen:2020lbd}, exceeding our upper limits by many orders of magnitude. Within the narrow log-normal benchmark considered here, our results therefore strongly disfavor PBHs as the progenitors of these observed mergers.

This conclusion becomes particularly acute for BHs in the mass gap, whose primordial interpretation has long been invoked to resolve the tension with standard stellar evolution but is now left with little room. As a result, these objects are redirected toward astrophysical formation channels. This inference is consistent with recent work \cite{Tong2026}, in which a mass gap is identified in the secondary black hole distribution, while primary black holes in this range can be explained as the products of hierarchical mergers. Our constraints therefore provide complementary evidence that astrophysical channels dominate the origin of these systems.



\begin{acknowledgments}

YCB would like to thank Chen Yuan for helpful discussions.
QGH is supported by the National Natural Science Foundation of China (Grant No.~12547110,~12475065, 12447101) and the China Manned Space Program with grant no. CMS-CSST-2025-A01. YMW is supported by the National Natural Science Foundation of China (Grant No.~12505086).
	
\end{acknowledgments}

\bibliography{./ref}


\onecolumngrid
\appendix
\newpage

\begin{center}
{\LARGE\textbf{SUPPLEMENTAL MATERIAL}}
\end{center}

\begin{center}
{\Large
Yan-Chen Bi,$^{1,2}$ Yu-Mei Wu,$^{3}$ and Qing-Guo Huang$^{1,2,4}$}
\end{center}

\begin{center}
{\large\textit{$^1$Institute of Theoretical Physics, Chinese Academy of Sciences (CAS), Beijing 100190, China}}
\end{center}
\begin{center}
{\large\textit{$^2$University of Chinese Academy of Sciences (UCAS), Beijing 100049, China}}
\end{center}
\begin{center}
{\large\textit{$^3$Center for Gravitation and Cosmology, College of Physical Science and Technology, Yangzhou University, Yangzhou, 225009, China}}
\end{center}
\begin{center}
{\large\textit{$^4$School of Fundamental Physics and Mathematical Sciences, Hangzhou Institute for Advanced Study (HIAS), University of Chinese Academy of Sciences (UCAS), Hangzhou 310024, China}}
\end{center}

\section{Likelihood function}\label{appa}
Here we present a more explicit derivation of the likelihood used in our analysis. 
We denote by $\mathbf{j}_{\mathrm{obs}}$ the observed data vector and by $\mathbf{j}_{\mathrm{GW}}$ the GW-induced signal,
both of length $N_p$, the number of pulsars. In the notation of \cite{DeRocco:2023qae}, the likelihood can be written in matrix form as
\m
\mathcal{L}(\mathbf{j}_{\mathrm{obs}} \mid \mathbf{j}_{\mathrm{GW}}) 
= \frac{1}{\sqrt{(2\pi)^{N_p} |\mathbf{N}|}} 
\exp\left[-\frac{1}{2} (\mathbf{j}_{\mathrm{obs}} - \mathbf{j}_{\mathrm{GW}})^T 
\mathbf{N}^{-1} (\mathbf{j}_{\mathrm{obs}} - \mathbf{j}_{\mathrm{GW}})\right].
\label{likelihood_original}
\n
Here, the noise covariance matrix is
$\mathbf{N}=\mathrm{diag}(\sigma_a^2)=\mathrm{diag}(\sigma_{\mathrm{obs}}^2+\sigma_{\mathrm{RN}}^2)$, reflecting the assumption that the measurement noise and ultra-low frequency red noise (RN) are uncorrelated between different pulsars. We then analytically marginalize over ${\mathbf{j}}_{\mathrm{GW}}$,
\m
\mathcal{L} = \int d\mathbf{j}_{\mathrm{GW}} \, 
\mathcal{L}(\mathbf{j}_{\mathrm{obs}} \mid \mathbf{j}_{\mathrm{GW}}) 
\, p(\mathbf{j}_{\mathrm{GW}}),
\n
assuming a Gaussian prior
\m
p({\mathbf{j}}_{\mathrm{GW}}) \propto \exp\[-\frac{1}{2} {\mathbf{j}}_{\mathrm{GW}}^T \mathbf{C}^{-1}_{\mathrm{GW}} {\mathbf{j}}_{\mathrm{GW}}\] .
\n
Here, $\mathbf{C}_{\mathrm{GW}}=\langle {\mathbf{j}}_{\mathrm{GW}} {\mathbf{j}}_{\mathrm{GW}}^T \rangle$ is the GW covariance matrix, and $\langle \cdot \rangle$ denotes the ensemble average. Defining $\mathbf{C}=\mathbf{N}+\mathbf{C}_{\mathrm{GW}}$, the integral can be carried out analytically, yielding
\m
\mathcal{L} = \frac{1}{\sqrt{(2\pi)^{N_p} |\mathbf{C}|}} 
\exp\left[-\frac{1}{2} \mathbf{j}_{\mathrm{obs}}^T 
\mathbf{C}^{-1} \mathbf{j}_{\mathrm{obs}}\right].
\n
We note that the Woodbury identity, $(\mathbf{N}+\mathbf{C}_{\mathrm{GW}})^{-1} = \mathbf{N}^{-1} - \mathbf{N}^{-1}(\mathbf{C}_{\mathrm{GW}}^{-1} + \mathbf{N}^{-1})^{-1}\mathbf{N}^{-1}$, is used to simplify the result. Compared with the numerical treatment adopted in \cite{DeRocco:2022irl,DeRocco:2023qae,Zheng:2025tcm}, this formulation provides a closed, fully analytic likelihood for the statistical analysis. These two approaches are physically equivalent. In particular, the GW contribution induces off-diagonal correlations between different pulsars through the covariance matrix in our approach, as given in \Eq{correlator}, whereas the noise contributes only to the diagonal entries.

\section{Dataset}\label{appb}
In \Table{tab:dataset}, we list the dataset of 46 pulsars used in this work. The dataset is publicly available in \cite{DeRocco:2022irl,DeRocco:2023qae}. The corresponding values of $\ddot{P}_{\mathrm{obs}}$ for each pulsar were computed in \cite{Liu:2019iuh}.

\begin{table*}[htbp]
\setlength{\tabcolsep}{10pt}
\centering
\caption{Pulsar dataset used for analysis. Here $l$ and $b$ denote the Galactic longitude and latitude, $d$ is the pulsar distance, $T$ is the observation timespan, and $\ddot{P}_{\mathrm{obs}} / P$ column is the measured jerk and measurement uncertainty $\sigma_{\mathrm{obs}}$. The quantity $\sigma_{\mathrm{RN}}$ represents the additional uncertainty from ultralow-frequency red noise, and Ref.\ indicates the source of the pulsar parameters. The dataset is publicly available in \cite{DeRocco:2022irl,DeRocco:2023qae}.}
\label{tab:dataset}
\begin{tabular}{lccccccc}
\hline\hline
Pulsar & $l$ (deg) & $b$ (deg) & $d$ (kpc) & $T$ (yr) & $\ddot{P}_{\rm obs}/P$ ($10^{-30}\,\mathrm{s^{-2}}$) & $\sigma_{\rm RN}$ ($10^{-30}\,\mathrm{s^{-2}}$) & Reference \\
\hline

J0030+0451 & 113.141 & -57.611 & 0.324 & 15.1 & $-4 \pm 4$ & 0.54 & \cite{EPTA:2016ndq} \\
J0034-0534 & 111.492 & -68.069 & 1.348 & 13.5 & $0 \pm 20$ & 0 & \cite{EPTA:2016ndq} \\
J0218+4232 & 139.508 & -17.527 & 3.150 & 17.6 & $-2 \pm 5$ & 0.14 & \cite{EPTA:2016ndq} \\
J0437-4715 & 253.394 & -41.963 & 0.157 & 14.9 & $-1 \pm 1$ & 0 & \cite{Reardon:2015kba} \\
J0610-2100 & 227.747 & -18.184 & 3.260 & 6.9 & $0 \pm 50$ & 0 & \cite{EPTA:2016ndq} \\
J0613-0200 & 210.413 & -9.305 & 0.780 & 16.1 & $0.6 \pm 0.6$ & 0.13 & \cite{EPTA:2016ndq} \\
J0621+1002 & 200.570 & -2.013 & 0.425 & 11.8 & $-70 \pm 30$ & 1.28 & \cite{EPTA:2016ndq} \\
J0711-6830 & 279.531 & -23.280 & 0.106 & 17.1 & $1 \pm 1$ & 0 & \cite{Reardon:2015kba} \\
J0751+1807 & 202.730 & 21.086 & 1.110 & 17.6 & $0 \pm 2$ & 0 & \cite{EPTA:2016ndq} \\
J0900-3144 & 256.162 & 9.486 & 0.890 & 6.9 & $-10 \pm 20$ & 0 & \cite{EPTA:2016ndq} \\
J1012+5307 & 160.347 & 50.858 & 0.700 & 16.8 & $0.4 \pm 0.7$ & 0.01 & \cite{EPTA:2016ndq} \\
J1022+1001 & 231.795 & 51.101 & 0.645 & 17.5 & $-2 \pm 1$ & 0.01 & \cite{EPTA:2016ndq} \\
J1045-4509 & 280.851 & 12.254 & 0.340 & 17.0 & $-2 \pm 7$ & 0.05 & \cite{Reardon:2015kba} \\
J1455-3330 & 330.722 & 22.562 & 0.684 & 9.2 & $6 \pm 20$ & 0.17 & \cite{EPTA:2016ndq} \\
J1600-3053 & 344.090 & 16.451 & 1.887 & 9.1 & $4 \pm 5$ & 0.06 & \cite{Reardon:2015kba} \\
J1603-7202 & 316.630 & -14.496 & 0.530 & 15.3 & $1 \pm 4$ & 0.02 & \cite{Reardon:2015kba} \\
J1640+2224 & 41.051 & 38.271 & 1.515 & 17.3 & $-0.9 \pm 0.9$ & 0 & \cite{EPTA:2016ndq} \\
J1643-1224 & 5.669 & 21.218 & 0.740 & 17.3 & $-2 \pm 2$ & 0.01 & \cite{EPTA:2016ndq} \\
J1713+0747 & 28.751 & 25.223 & 1.311 & 17.7 & $-0.5 \pm 0.5$ & 0.38 & \cite{EPTA:2016ndq} \\
J1721-2457 & 0.387 & 6.751 & 1.393 & 12.7 & $-30 \pm 70$ & 0.01 & \cite{EPTA:2016ndq} \\
J1730-2304 & 3.137 & 6.023 & 0.620 & 16.9 & $0 \pm 2$ & 0 & \cite{Reardon:2015kba} \\
J1732-5049 & 340.029 & -9.454 & 1.873 & 8.0 & $20 \pm 20$ & 0.03 & \cite{Reardon:2015kba} \\
J1738+0333 & 27.721 & 17.742 & 1.471 & 7.3 & $-30 \pm 90$ & 0 & \cite{EPTA:2016ndq} \\
J1744-1134 & 14.794 & 9.180 & 0.395 & 17.3 & $0.8 \pm 0.8$ & 0.02 & \cite{EPTA:2016ndq} \\
J1751-2857 & 0.646 & -1.124 & 1.087 & 8.3 & $-10 \pm 50$ & 0 & \cite{EPTA:2016ndq} \\
J1801-1417 & 14.546 & 4.162 & 1.105 & 7.1 & $-30 \pm 100$ & 0.02 & \cite{EPTA:2016ndq} \\
J1802-2124 & 8.382 & 0.611 & 0.760 & 7.2 & $10 \pm 60$ & 0.01 & \cite{EPTA:2016ndq} \\
J1804-2717 & 3.505 & -2.736 & 0.805 & 8.1 & $-40 \pm 40$ & 0 & \cite{EPTA:2016ndq} \\
J1843-1113 & 22.055 & -3.397 & 1.260 & 10.1 & $-7 \pm 20$ & 0.05 & \cite{EPTA:2016ndq} \\
J1853+1303 & 44.875 & 5.367 & 2.083 & 8.4 & $-30 \pm 20$ & 0 & \cite{EPTA:2016ndq} \\
B1855+09   & 42.290 & 3.060 & 1.200 & 17.3 & $1 \pm 2$ & 0.03 & \cite{EPTA:2016ndq} \\
J1909-3744 & 359.731 & -19.596 & 1.140 & 9.4 & $0.6 \pm 0.9$ & 0.02 & \cite{EPTA:2016ndq} \\
J1910+1256 & 46.564 & 1.795 & 1.496 & 8.5 & $30 \pm 20$ & 0 & \cite{EPTA:2016ndq} \\
J1911+1347 & 25.137 & -9.579 & 1.069 & 7.5 & $14 \pm 8$ & 0 & \cite{EPTA:2016ndq} \\
J1911-1114 & 47.518 & 1.809 & 1.365 & 8.8 & $20 \pm 50$ & 0 & \cite{EPTA:2016ndq} \\
J1918-0642 & 30.027 & -9.123 & 1.111 & 12.8 & $0 \pm 8$ & 2.46 & \cite{EPTA:2016ndq} \\
B1953+29   & 65.839 & 0.443 & 6.304 & 8.1 & $-20 \pm 50$ & 0 & \cite{EPTA:2016ndq} \\
J2010-1323 & 29.446 & -23.540 & 2.439 & 7.4 & $20 \pm 20$ & 0 & \cite{EPTA:2016ndq} \\
J2019+2425 & 64.746 & -6.624 & 1.163 & 9.1 & $-500 \pm 900$ & 0 & \cite{EPTA:2016ndq} \\
J2033+1734 & 60.857 & -13.154 & 1.740 & 7.9 & $40 \pm 100$ & 0 & \cite{EPTA:2016ndq} \\
J2124-3358 & 10.925 & -45.438 & 0.410 & 16.8 & $0 \pm 3$ & 0.02 & \cite{Reardon:2015kba} \\
J2129-5721 & 338.005 & -43.570 & 3.200 & 15.4 & $-1 \pm 2$ & 0 & \cite{Reardon:2015kba} \\
J2145-0750 & 47.777 & -42.084 & 0.714 & 17.5 & $-2 \pm 1$ & 0.28 & \cite{EPTA:2016ndq} \\
J2229+2643 & 87.693 & -26.284 & 1.800 & 8.2 & $-20 \pm 20$ & 0 & \cite{EPTA:2016ndq} \\
J2317+1439 & 91.361 & -42.360 & 1.667 & 17.3 & $-1 \pm 3$ & 0 & \cite{EPTA:2016ndq} \\
J2322+2057 & 96.515 & -37.310 & 1.011 & 7.9 & $30 \pm 70$ & 0 & \cite{EPTA:2016ndq} \\

\hline\hline
\end{tabular}
\end{table*}

\end{document}